\documentclass[preprint,showpacs,preprintnumbers,amsmath,amssymb,nofootinbib,showkeys,aps]{revtex4}
\usepackage{graphicx}% Include figure files
\usepackage{dcolumn}% Align table columns on decimal point
\usepackage{bm}% bold math
\usepackage{latexsym}
\usepackage{epsfig}
\usepackage{color}
\usepackage{bm}% bold math

\newcommand{\be}{\begin{equation}}
\newcommand{\ee}{\end{equation}}
\newcommand{\beqq}{\setlength\arraycolsep{2pt}\begin{eqnarray}}
\newcommand{\eeqq}{\vspace{0cm} \end{eqnarray}}
\newcommand{\bea}{\begin{eqnarray}}
\newcommand{\eea}{\end{eqnarray}}

\begin{document}

\title{CCDM Model with Spatial Curvature and The Breaking of ``Dark Degeneracy''}

\author{J. F. Jesus} \email{jfjesus@itapeva.unesp.br}
\affiliation{Universidade Estadual Paulista ``J\'ulio de Mesquita Filho'' -- Campus Itapeva \\
Rua Geraldo Alckmin 519, 18409-010, Vila N. Sra. de F\'atima, Itapeva, SP, Brazil}

\author{F. Andrade-Oliveira}\email{foliveira@astro.iag.usp.br}
\affiliation{Departamento de Astronomia, Universidade de S\~{a}o Paulo,\\ Rua do Mat\~{a}o, 1226,\\ 05508-900, S\~{a}o Paulo, SP, Brazil}

%\keywords{dark matter theory, particle physics - cosmology connection }

\begin{abstract}
Creation of Cold Dark Matter (CCDM), in the context of Einstein Field Equations, leads to a negative creation pressure, which can be used to explain the accelerated expansion of the Universe. Recently, it has been shown that the dynamics of expansion of such models can not be distinguished from the concordance $\Lambda$CDM model, even at higher orders in the evolution of density perturbations, leading at the so called ``dark degeneracy''. However, depending on the form of the CDM creation rate, the inclusion of spatial curvature leads to a different behavior of CCDM when compared to $\Lambda$CDM, even at background level. With a simple form for the creation rate, namely, $\Gamma\propto\frac{1}{H}$, we show that this model can be distinguished from $\Lambda$CDM, provided the Universe has some amount of spatial curvature. 
Observationally, however, the current limits on spatial flatness from CMB indicate that neither of the models are significantly favored against the other by current data, at least in the background level.
\end{abstract}

%
%\flushbottom
%\pacs{}
%\keywords{}
\maketitle

%====================================================================================================================
%====================================================================================================================

\section{\label{introduction} Introduction}

Since 1998, the series of observations of the luminosity-redshift relation of the Supernovae Type Ia (SNe Ia) has set a new era in cosmology \cite{riess1997,perlmutter,riess07,union08}. Those observations, complemented by the observations of the Cosmic Microwave Background (CMB) anisotropies, Baryon Acoustic Oscillations (BAO), Hubble parameter in different redshifts,
%  inter alia,
strongly suggest that the Universe has a great order of spatial flatness and has entered in a late phase of accelerating expansion \cite{riess07,union08,union21, planck, bao, hz}.

Inside the context of relativistic cosmology, the accelerated expansion of the Universe is usually attributed to a new dark component called dark energy which possesses as main feature the negative pressure. The most favoured candidate for dark energy is the cosmological constant, $\Lambda$. This new component not only can fit the SN Ia observations but can also recover the flatness of the spatial hypersection of the Universe, as predicted by inflationary theory and corroborated by  CMB observations.

Nevertheless, this concordance model has some serious issues. For example, the density of cosmological constant $\Lambda$, responsible for the late acceleration of expansion of the Universe as estimated from observational data ($\rho_\Lambda$) is almost null ($10^{-47}$ g/cm$^3$) and must be fine tuned to explain quantitatively the acceleration. Due to its equation of state ($p_\Lambda=-\rho_\Lambda$), the cosmological constant could rise from the vacuum energy of quantum fields. However, the theoretical estimation to the vacuum energy through quantum field theory ($\rho_{v,Th}$) is up to 122 orders of magnitude bigger than the observational density \cite{weinberg89}.

Inside the General Relativity, the negative pressure is the key element for acceleration. This can occur naturally in thermodynamical process departing from the equilibrium, for instance,
the matter creation process at expenses of gravity. This phenomenon gives rise to  a term of negative pressure  which should be considered at the level of Einstein Equations, as  shown by Prigogine and collaborators and formulated in a manifestly covariant way by Lima, Calv\~{a}o \& Waga \cite{prigogine89,CLW}. The inclusion of the backreaction at the level of Einstein Field Equations was determinant to the rise of a new class of cosmological models with matter creation.

Many CCDM models have been proposed in the literature, each of those have different phenomenological creation rate dependencies. Among them, recently, a model from Lima, Jesus and Oliveira (LJO) was interesting because it was shown that this model leads to the same background evolution as the concordance $\Lambda$CDM model, for any spatial curvature. Further, it was shown that even at linear density perturbation evolution, such a degeneracy persisted, leading to the so called ``dark degeneracy''.

A first alternative to break such a degeneracy was given by Jesus and Pereira \cite{jesuspereira}, which have found, directly from quantum field calculations, a creation rate which depended on the dark matter mass. They have used observational data to constrain the dark matter mass and have argued that with better constraints in the future, this could be used to distinguish CCDM from $\Lambda$CDM.

In this paper we investigate a cosmological model driven only by cold dark matter (CDM) creation at expenses of gravitational field in which the rate of CDM creation evolves reciprocally to the expansion rate, and we include the possibility of nonzero spatial curvature. We assume that the created particles are described by a real scalar field and consequently the created particles are its own antiparticles. 
 Similarly to the standard model, the scenario presented here has the same degree of freedom and is also capable of explaining the accelerating expansion. We show that, with some amount of spatial curvature, the CCDM behavior differs from $\Lambda$CDM and both can be distinguished using observational data.

In Section II,
we discuss the dynamics of the universe with the pressure due to creation.  In
Section III, we discuss an specific rate of dark matter creation. In Section IV,  we constrain the free parameters of the model. In Section V, we compare the model with other models on the literature. Finally, we summarize the main results in conclusion.

%====================================================================================================================
%====================================================================================================================

\section{\label{dinamical}  Cosmic Dynamics on Models with Creation of CDM Particles }

We will start by considering the homogeneous and isotropic FRW line element (with $c=1$):

\begin{equation}
\label{ds2}
  ds^2 = dt^2 - a^{2}(t) \left(\frac{dr^2}{1-k r^2} + r^2 d\theta^2+
      r^2{\rm sin}^{2}\theta d \phi^2\right),
\end{equation}
\noindent where $k$ can assume values $-1$, $+1$ or $0$.

In this background, the Einstein field equations are given by 

\begin{equation}
\label{fried1}
    8\pi G (\rho_{rad} + \rho_{b}  + \rho_{dm}) = 3 \frac{\dot{a}^2}{a^2} + 3 \frac{k}{a^2},
\end{equation}

\noindent and

\begin{equation}
\label{fried2}
   8\pi G (p_{rad} + p_{c}) =  -2 \frac{\ddot{a}}{a} - \frac{\dot{a}^2}{a^2} -
	\frac{k}{a^2}.
\end{equation}
where $\rho_{rad}$, $\rho_{b}$ and $\rho_{dm}$ are  the density parameters of radiation, baryons and dark matter,  $p_{rad}=\rho_{rad}/3$ is the radiation pressure and $p_c$ is the creation pressure.

For the radiation and baryon components, the energy conservation  are given by the usual expressions:

\begin{equation}\label{rad}
\dot{\rho}_{rad} +  4 \frac{\dot{a}}{a}\rho_{rad} = 0,
\end{equation}
and
\be\label{consbar}
\dot{\rho}_{b} + 3 \frac{\dot{a}}{a}{\rho}_{b} = 0.
\ee
where each overdot means one time derivative and we have used that $p_{rad}=\rho_{rad}/3$ and $p_{b}=0$.

On the other hand, when the creation process is considered we should take into account a matter creation source at level of Einstein Field Equations \cite{CLW}:

\begin{equation}
\label{ConsDM}
\frac{\dot{\rho}_{dm}}{{\rho}_{dm}} + 3 \frac{\dot{a}}{a} = \Gamma \rho_{dm},
\end{equation}
where $\Gamma$ is the rate of dark matter creation in units of (time)${}^{-1}$.

As shown by \cite{CLW}, the creation of dark matter is responsible for an extra pressure term at the level of Einstein Equations, the so called creation  pressure, $p_c$,
 
\begin{equation}
\label{pc}
  p_{c} = -\frac{\rho_{dm} \Gamma}{3H},
\end{equation}
where we have considered an ``adiabatic" creation, i.e., the case when the entropy per particle is constant.

As a consequence of the above equation, one can see that the dynamics of the universe is directly affected by the rate of creation of cold dark matter, $\Gamma$. In particular, in the case $\Gamma > 0$ (creation of particles) we have a negative pressure creation and in the case $\Gamma \to 0$  we recover the well known dynamics when the universe is lately dominated by pressureless matter (baryons plus dark matter).

%====================================================================================================================
%====================================================================================================================
\section{\label{model} Creation of Cold Dark Matter (CCDM) Model}

The difficulties in identifying the nature of dark energy led the cosmologists to a quest for better candidates to explain the late acceleration of the Universe.
In  the literature, models with CDM creation has been discussed as a viable explanation to this recent phenomenon. It has been shown that under a convenient choice of the particle creation rate $\Gamma$, this scenario is able to support the observed  dynamics,  linear structure formation and thermodynamics features of the late Universe \cite{LSS08,ljo10,jobl,CLW} . 

We argue that a natural dependence of the CDM creation rate is on the expansion rate, as already proposed in other CCDM models present on the literature. It has already been studied a linear dependence with the Hubble parameter \cite{LSS08}, and at the time of writing this paper, it has been proposed a power law dependence on Hubble parameter \cite{graef}. It is clear to us that the expansion acceleration is a recent feature, so the CDM creation also must be recent. As the Universe evolves, the Hubble parameter decreases, and $\Gamma$ must increase. We can satisfy those conditions by assuming $\Gamma$  to be a negative power law of $H$, or, in the simplest case, $\Gamma\propto H^{-1}$.
So, in this work, we consider the following creation rate:

\begin{equation}
\label{Gamma} 
\Gamma=3\alpha \frac{H_0^2}{H},%+3\beta H
\end{equation}
where $\alpha$ is a constant free parameter of the model which drives the creation rate and the factor $3H_0^2$ has been introduced for mathematical convenience. This is also interesting because, as we shall see, with the identification $\alpha=\Omega_{\Lambda}$, the flat $\Lambda$CDM is a particular case of our CCDM model, when we neglect the baryon contribution.

Since we are considering only the late phase of the dynamics of the universe, we can neglect the radiation terms from now on. Thus, by combining Eqs. (\ref{fried1}) and (\ref{fried2}), we have
\begin{equation}
 \frac{\ddot{a}}{a}=-\frac{4\pi G}{3}(\rho_{b} + \rho_{dm} +3p_c)
\end{equation}
Replacing $p_c$ from Eq. (\ref{pc}), we may write
\begin{equation}
 \frac{\ddot{a}}{a}=-\frac{4\pi G}{3}\left[\rho_{b} + \rho_{dm}\left(1-\frac{\Gamma}{H}\right)\right]
\end{equation}
Using that $\frac{\ddot{a}}{a}=\dot{H}+H^2$ and changing variables from time to redshift, we find
\begin{equation}
\frac{dH}{dz}=\frac{H}{1+z}+\frac{H_0^2\Omega_b(1+z)^2\Gamma}{2H^2}+\frac{H^2-H_0^2\Omega_k(1+z)^2}{2H(1+z)}\left(1-\frac{\Gamma}{H}\right)
\end{equation}
where we have used the solution of (\ref{consbar}) to baryon density, $\rho_b=\rho_{b0}(1+z)^3$, $\Omega_b=\frac{\rho_{b0}}{\rho_{c0}}$ is the present baryon density parameter, and $\Omega_k=-\frac{k}{H_0^2}$ is the present curvature density parameter. Replacing the creation rate (\ref{Gamma}) and changing to dimensionless variable $y\equiv\left(\frac{H}{H_0}\right)^2$, we find
\begin{equation}
\frac{dy}{dz}=3\frac{y-\alpha}{1+z}+\frac{3\alpha\Omega_b(1+z)^2}{y}-\Omega_k(1+z)\left(1-\frac{3\alpha}{y}\right)
\label{odey}
\end{equation}
which can not be solved analytically, except for some special cases. For example, if the Universe is spatially flat and we can neglect baryons ($\Omega_k=\Omega_b=0$), we can solve (\ref{odey}) to find:
%there is no baryons nor spatial curvature 
\begin{equation}
H(z)^2=H_0^2[\alpha+(1-\alpha)(1+z)^3]
\end{equation}
which corresponds to the flat CCDM LJO model, or to flat $\Lambda$CDM model, with the identification $\alpha=\Omega_\Lambda$. However, by introducing baryons or curvature, this model can not recover the LJO or the $\Lambda$CDM model anymore. So, this model can be useful to discriminate between CCDM models and $\Lambda$CDM model even in the background level. On the other hand, the LJO  and $\Lambda$CDM models can only be distinguished at the perturbation levels \cite{jobl} and only in the absence of the  separation of dark matter components  \cite{RamosEtAl14}.

There is also one more analytical solution. If we neglect curvature ($\Omega_k=0$), retaining baryon density ($\Omega_b\neq0$), we find an analytical solution to (\ref{odey}),

\begin{equation}
\left(\frac{H}{H_0}\right)^2=\Omega _b(1+z)^3 \left[1+W\left(\frac{\left(1-\Omega _b\right) e^{\frac{1-\alpha }{\Omega _b}+\frac{\alpha
   }{(1+z)^3 \Omega _b}-1}}{\Omega _b}\right)\right]
	\label{Hzwb}
\end{equation}
where $W(x)$ is the principal Lambert W function, also known as product logarithm, real solution to equation $x=W(x)e^{W(x)}$.

However, if we neglect baryons only ($\Omega_b=0$, $\Omega_k\neq0$), or if we consider the full equation (\ref{odey}), with $\Omega_b\neq0$ and $\Omega_k\neq0$, we can not find an analytical solution to $H(z)$, and we have to resort to numerical methods. Nevertheless, if curvature and baryonic contributions can both be considered small ($0<\Omega_b\ll1$, $|\Omega_k|\ll 1$), we can find an approximation,
\begin{eqnarray}
\left(\frac{H}{H_0}\right)^2&=&\alpha+(1-\alpha)(1+z)^3+\frac{\Omega_k}{2}\left[1-(1+z)^2\right]+3\Omega_k\log(1+z)+\nonumber\\
&+&\left(\frac{\Omega_b\alpha}{1-\alpha}-\Omega_k\right)\log\left[\alpha+(1-\alpha)(1+z)^3\right]
	\label{HzAprox}
\end{eqnarray}
which we have found by solving (\ref{odey}) with $\Omega_b=\Omega_k=0$, replacing the solution on (\ref{odey}) and solving again, with $\Omega_b\neq0$, $\Omega_k\neq0$.

\begin{figure}[t]
\centerline{%\epsfig{figure=wk00alpha00_07_09GRB.eps,width=0.45\linewidth,angle=0}
 \hspace{0.08\linewidth}
\epsfig{figure=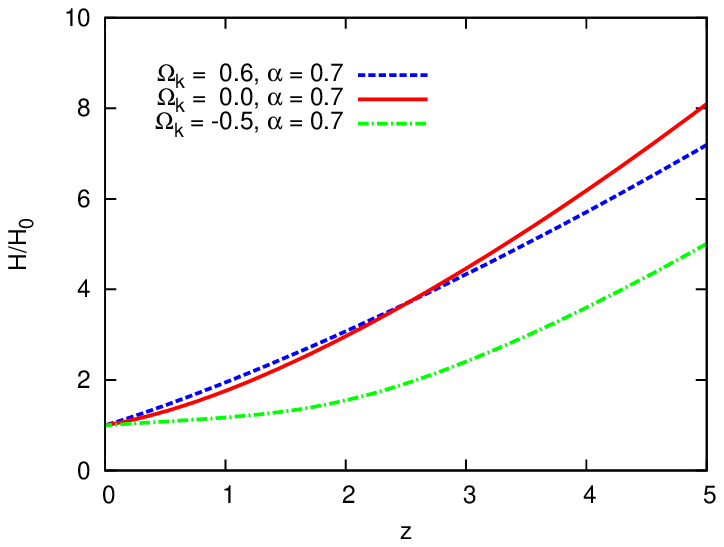,width=0.48\linewidth,angle=0}
\hspace{0.0\linewidth}
\epsfig{figure=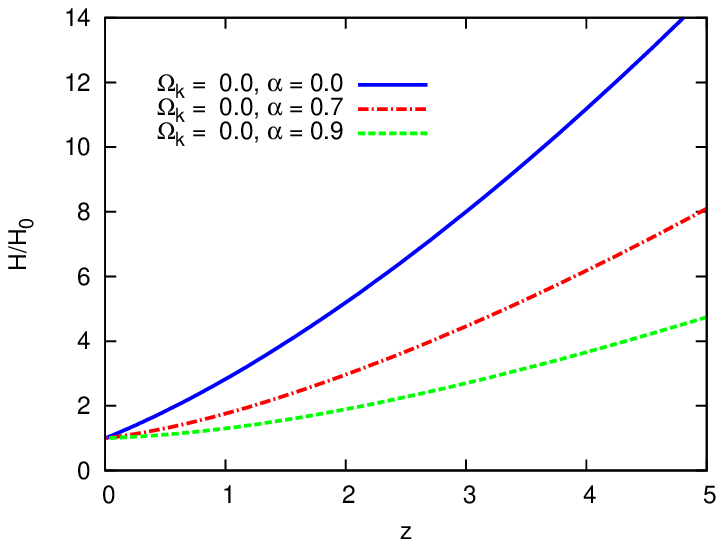,width=0.48\linewidth,angle=0}
}%-90}}height=3.8truein,
\caption{Numerical solutions for $H/H_0$  in function of the redshift $z$, and its sensitivity to the free parameters $\alpha$ and $\Omega_k$. {\bf Left)}  Evolution of $H(z)$ with $\alpha=0.7$ and $\Omega_k=0$ (solid curve), $\Omega_k=0.6$ (dashed curve) and  $\Omega_k=-0.5$ (dashed-dotted curve). 
 {\bf Right)} Evolution of $H(z)$ with $\Omega_k=0$ and $\alpha= 0.0$ (solid curve), $\alpha= 0.7$ (dashed-dotted curve) and  $\alpha= 0.9$ (dashed curve).
}
\label{fighz}
\end{figure}

In Figure \ref{fighz} we show the numerical solutions of $H(z)/H_0$ for some values of the free parameters of the model, namely, $\Omega_k$ and $\alpha$. It is worthy to remark that the model has as particular cases two well known models: the Einstein-de Sitter, for $\Omega_k = 0$ and $\alpha=0$ and the flat $\Lambda$CDM for  $\Omega_k=0$ and $\alpha\sim0.7$ and $\Omega_b=0$.

%====================================================================================================================
%====================================================================================================================
\section{\label{observational} Observational Constraints}
 In this section, we obtain constraints to the free parameters of the model, namely,  $\Omega_{k}$ and $\alpha$. In order to do this, we considered some measurements of the Hubble parameter, $H(z)$ \cite{farooq2013} and the Supernovae Type Ia (SN Ia) dataset of Union 2.1 \cite{union21}.

\subsection{$H(z)$ Constraints}

Hubble parameter data as function of redshift yields one of the most straightforward constraints on cosmological models. In general, these observational data depend on astrophysical assumptions, not depending on any background cosmological model. However, is not easy to obtain such data, as it is not directly observable, but is rather inferred from astrophysical observations.

At the present time, the most important methods for obtaining $H(z)$ data are
(i)  through  ``cosmic chronometers", for example, the differential age of galaxies (DAG), (ii) measurements of peaks of acoustic oscillations of baryons (BAO)  and (iii) through correlation function of luminous red galaxies (LRG) \cite{hz}.
In this work, we use the data compilation of $H(z)$ from Farooq and Ratra \cite{farooq2013}, which is, currently, the most complete compilation, with 28 measurements.

From these data, we perform a $\chi^2$-statistics, generating the $\chi^2_H$ function of free parameters:
\begin{equation}
 \chi^2_H=\sum_{i=1}^{28}\left[\frac{H_0E(z_i,\alpha,\Omega_k,\Omega_b)-H_i}{\sigma_{Hi}}\right]^2
\end{equation}
where $E(z)\equiv\frac{H(z)}{H_0}$ and $H(z)$ is obtained by solving numerically Eq. (\ref{odey}).

Throughout every analysis on this paper, we fix the baryon density parameter at the value estimated by Planck and WMAP: $\Omega_b=0.049$ \cite{planck}, a value which is in agreement with Big Bang Nucleosynthesis (BBN), as shown on Ref. \cite{pdg}.

As the function to be fitted, $H(z)=H_0E(z)$, is linear on the Hubble constant, $H_0$, we may analitically project over $H_0$, yielding $\tilde{\chi}^2_H$:
\begin{equation}
\tilde{\chi}^2_{H}=C-\frac{B^2}{A}
\end{equation}
where $A\equiv\sum_{i=1}^{n}\frac{E_i^2}{\sigma_{Hi}^2}$, $B\equiv\sum_{i=1}^n\frac{E_iH_i}{\sigma_{Hi}^2}$, $C\equiv\sum_{i=1}^n\frac{H_i^2}{\sigma_{Hi}^2}$ and $E_i\equiv\frac{H(z_i)}{H_0}$. The result of such analysis can be seen on Figure \ref{contours} (left). As can be seen, the results from $H(z)$ data alone yield very loose constraints on the plane $\Omega_k$-$\alpha$. In fact, over the region $0<\alpha<1.4$ and $-1<\Omega_k<1$, only the 68\% c.l. statistical contour could close. The minimum $\chi^2$ was $\chi^2_{min}=16.269$, which is too low for 28 data, yielding a $\chi^2$ per degree of freedom $\chi^2_\nu=0.626$. The best fit parameters were $\alpha=0.791^{+0.18}_{-0.085}$,
 $\Omega_k=0.04^{+0.46}_{-0.40}$ in the joint analysis. We believe that such a loose constraint can be due to a underestimate of uncertainties on the $H(z)$ compilation data, which is evidenced by its low $\chi^2_\nu$. This issue has been addressed by \cite{farooq2013} by combining $H(z)$ data with other constraints. While \cite{farooq2013} uses a prior over $H_0$, we choose to use a prior over $\Omega_k$, as a prior over $H_0$ did not affect much the constraints found with $H(z)$ data only. We have considered a prior over $\Omega_k$ based on Planck and WMAP results \cite{planck}.

Planck + WMAP indicate $\Omega_k=-0.037^{+0.043}_{-0.049}$, at 95\% c.l., in the context of $\Lambda$CDM.  Based on this result, along with the symmetrization process suggested by \cite{dagostini}, we use a prior of $\Omega_k\pm\sigma_{\omega_k}=-0.043\pm0.046$. We will refer to this simply as the CMB prior. The results can be seen on Figure \ref{contours} (right). As can be seen there, the limits over $\alpha$ and $\Omega_k$ are quite better. We have found $\chi^2_{min}=16.344$, $\alpha=0.775^{+0.059+0.098}_{-0.064-0.11}$$^{+0.14}_{-0.16}$ and $\Omega_k=-0.041\pm0.069\pm0.11\pm0.16$. In order to improve these constraints, we made a combined analysis with SN Ia data.

\subsection{Supernovae Type Ia  Bounds}

The parameters dependent distance modulus for a supernova at the redshift $z$ can be computed through the expression
\be
\label{mu}
\mu(z|\mathbf{s}) = m - M = 5\log d_L + 25,
\ee
where $m$ and $M$ are respectively the apparent and absolute magnitudes,  $\mathbf{s}\equiv (H_0, \alpha, \Omega_k)$ is the set of the free parameters of the model and $d_L$ is the luminosity distance in unit of Megaparsecs.

Since in the general case $H(z)$ has not an analytic expression, we must define $d_L$ through a differential equation. The luminosity distance $d_L$ can be written in terms of a dimensionless comoving distance $D$ by:
\begin{equation}
d_L=(1+z)\frac{H_0}{c}D
\end{equation}

The comoving distance can be related to $H(z)$, taking into account spatial curvature, by the following relation \cite{clarkson}:
\be
\label{clark}
\left(\frac{H}{H_0}\right)^2\equiv y=\frac{\Omega_{k}D^2+1}{D'^2},
\ee
where the prime denotes derivation with respect to redshift $z$. Inserting this relation in Eq. (\ref{odey}), we obtain a differential equation for $D$:
\begin{eqnarray}
D''&=& \frac{\Omega_kDD'^2}{1+\Omega_kD^2} - \frac{3D'}{2(1+z)}+\frac{D'^3}{2(1+\Omega_kD^2)}\left[\frac{3\alpha}{1+z}+\Omega_k(1+z)\left(1-\frac{3\alpha D'^2}{1+\Omega_kD^2}\right)\right]-\nonumber\\
&-&\frac{3\alpha\Omega_b(1+z)^2D'^5}{2(1+\Omega_kD^2)^2},
\end{eqnarray}

To solve numerically this equation we have used  as initial conditions $D(0)=0$ and $D'(0)=1$. In order to constrain the free parameters of the model we considered the Union 2.1 SN Ia dataset from Suzuki et al. \cite{union21}. The best-fit  set of parameters $\mathbf{s}$ was estimated from a $\chi^2$ statistics with %Amanullah =union 2.0

\be
\chi^2_{SN}=\sum^{N}_{i=1}\frac{\left[ \mu^i(z|\mathbf{s})-\mu^i_o(z)\right]^2}{\sigma_i^2}
\ee
where $\mu^i(z|\mathbf{s})$ is given by (\ref{mu}), $\mu^i_o(z)$ is the corrected distance modulus for a given SNe Ia at $z_i$ being $\sigma_i$ its corresponding individual uncertainty and $N=580$ for the Union 2.1 data compilation.
 
\begin{figure}[t]
\centerline{
\epsfig{figure=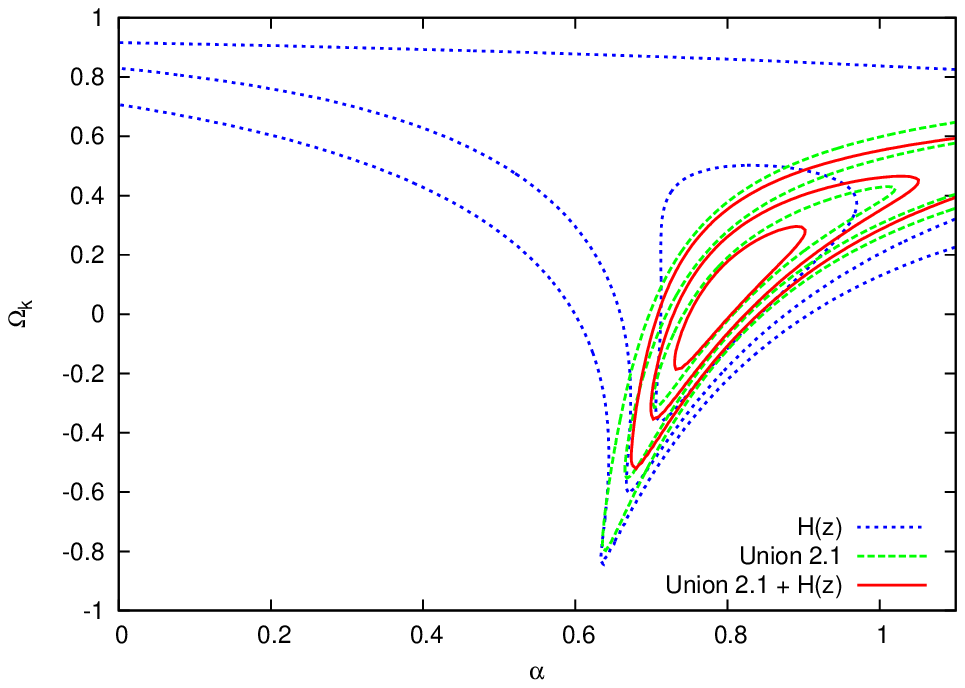,width=8cm}
\epsfig{figure=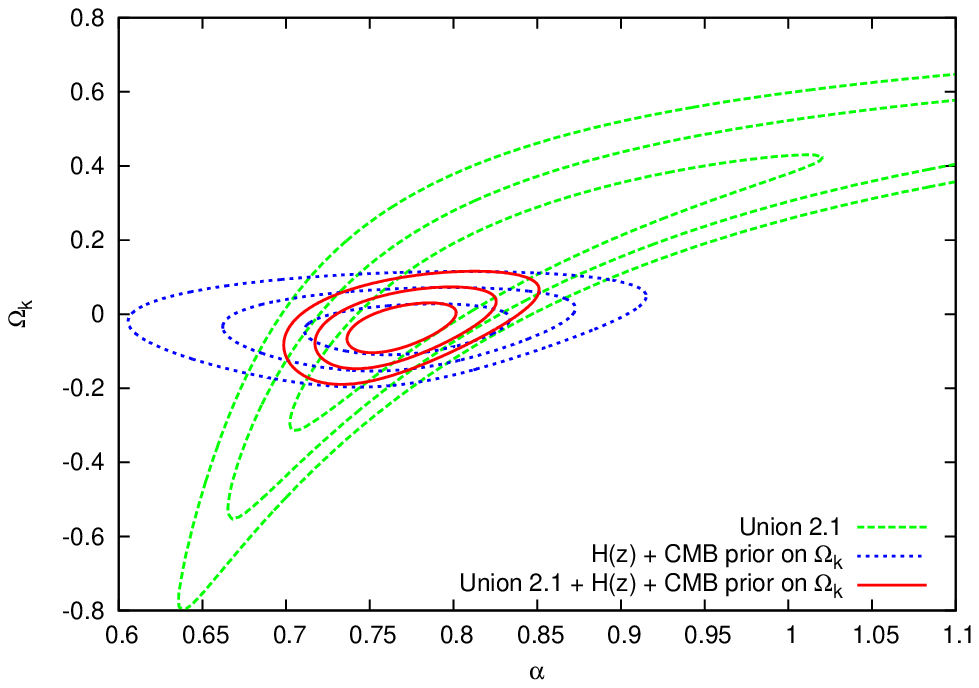,width=8cm}
}%-90}}height=3.8truein,
\caption{The results of our statistical analysis, with contours for 68.3\%, 95.4\% and 99.7\% confidence intervals. In both panels, the dashed lines correspond to SNs Union 2.1 data. {\bf Left)} Dotted lines: constraints from $H(z)$ data, solid lines: constraints from SNs + $H(z)$. {\bf Right)} Solid lines: constraints from SNs + $H(z)$ + CMB prior on $\Omega_k$.}
\label{contours}
\end{figure}

In Figure \ref{contours} (left), we display the space of parameters $\Omega_{k} -\alpha$ and the contours for $1\sigma$, $2\sigma$ and $3\sigma$ of confidence intervals. Using SN data alone, and marginalizing the nuisance parameter $h$ ($H_0=100h$ km s$^{-1}$Mpc$^{-1}$), we constrain the free parameters as $\alpha=0.788_{-0.086}^{+0.23}$, $\Omega_{k} =0.04_{-0.35}^ {+0.39}$ at 68\% confidence level and  $\chi^2_{\nu}=0.974$. We also show, on Figure \ref{contours} (left), the combination SNs Ia + $H(z)$, obtained by adding $\chi^2_{SN}+\chi^2_H$ and marginalizing over $h$. As can be seen, adding $H(z)$ data alone to SN Ia data, barely changes the SN constraints. In fact, we have found, on this case, $\alpha=0.792_{-0.061-0.093}^{+0.11\,\,\,+0.26}$, $\Omega_k=0.05_{-0.24-0.40}^{+0.25+0.42}$, with $\chi^2_\nu=0.956$. The main difference on this case, is that the 95\% c.l. contours could close on the region considered.

So, in order to find the best possible constraints with the data available, we made the full combination of SNs + $H(z)$ + CMB prior on $\Omega_k$. In this case, the constraints were quite restrictive, as shown in Figure \ref{contours} (right). As one can see, the CMB prior makes a great cut over the Union 2.1 contours, as $\Omega_k$ is strongly constrained, in this case. We have found, on this case, in the joint analysis: $\alpha=0.768_{-0.031-0.051-0.069}^{+0.034+0.058+0.084}$, $\Omega_k=-0.036_{-0.068-0.11-0.15}^{+0.066+0.11+0.15}$, $\chi^2_\nu=0.957$. Table \ref{tab1} summarizes the analysis results, including the analysis with the $H_0$ prior, based on the current limit on $H_0$ given by \cite{Humphreys13}, $H_0=72.0\pm3.0$ km s$^{-1}$Mpc$^{-1}$.

\newpage

\begin{table}[h]
\centering
\renewcommand{\arraystretch}{1.2}
\setlength{\tabcolsep}{5pt}
  \begin{tabular}{ |c | c | c| c|}
	    \hline
    Data & $\alpha$ & $\Omega_k$ &$\chi^2_\nu$\\ \hline\hline
    $H(z)$ & $0.791_{-0.085}^{+0.18}$ & $0.04_{-0.40}^{+0.46}$ & 0.626\\ \hline
    SNs & $0.788_{-0.086}^{+0.23}$ & $0.04_{-0.35}^{+0.39}$ & 0.973\\    \hline
		SNs$+H(z)$ & $0.792_{-0.061-0.093}^{+0.11\,\,\,+0.26}$ & $0.05_{-0.24-0.40}^{+0.25+0.42}$ & 0.955\\ \hline
		SNs$+H(z)+H_0$ prior & $0.787^{+0.11\,\,\,+0.25}_{-0.065-0.096}$ & $0.01^{+0.26+0.43}_{-0.26-0.42}$ & 0.957\\ \hline
		$H(z)+H_0$ prior & $0.786^{+0.17+0.61}_{-0.085-0.12}$ & $-0.06^{+0.35+0.61}_{-0.35-0.57}$ & 0.649\\ \hline
		$H(z)+$CMB prior & $0.775_{-0.064-0.11\,\,\,-0.17}^{+0.059+0.098+0.14}$ & $-0.041\pm0.069\pm0.11\pm0.16$ & 0.629\\ \hline
		SNs$+H(z)+$CMB prior & $0.768_{-0.031-0.051-0.069}^{+0.034+0.058+0.084}$ & $-0.036_{-0.068-0.11-0.15}^{+0.066+0.11+0.15}$ & 0.955\\
		\hline
  \end{tabular}
			\caption{Results of the joint analysis for the different combinations of data. Limits on the parameters correspond to 68.3\%, 95.4\% and 99.7\% c.l. as explained on text.}
			\label{tab1}
\end{table}

\section{Comparison with CCDM and $\Lambda$CDM models}
In the absence of baryons and if the spatial curvature vanishes, this model coincides both with the concordance $\Lambda$CDM model and LJO model \cite{ljo10}, where $\Gamma\propto\frac{H}{\rho}$. Also, this model has the same %is an special case of the recently proposed CCDM model, where $\Gamma\propto H^{-n}$, with $n=1$.
creation rate as the CCDM1 model of Ref. \cite{GraefEtAl14}. However, in their analysis, they have not included baryons nor considered nonzero spatial curvature. As baryons are separately conserved, the DM creation rate $\Gamma\propto\frac{1}{H}$ leads to a huge difference on the Universe evolution, as we have shown on our analytical solution, Eq. (\ref{Hzwb}). Numerically, taking baryons into account, for $\Omega_b\sim0.04$, leads to a relative difference of about 10$\%$ on $H(z)$, for $z\gtrsim 1$.

It has already been shown that the LJO model can not be distinguished from $\Lambda$CDM, for  any value of spatial curvature, neither at background level \cite{ljo10} nor at perturbation level \cite{RamosEtAl14}. In this manner, LJO gives rise to the so called ``dark degeneracy'' \cite{RamosEtAl14}, where, through cosmological observations, one can not determine if it is the quantum vacuum energy contribution ($\Lambda$CDM) or the quantum vacuum dark matter creation (CCDM) which accelerates the Universe. However, in the model proposed here, if the Universe has some amount of spatial curvature, we can distinguish it from $\Lambda$CDM.

\begin{figure}[t]
\centerline{%\epsfig{figure=wk00alpha00_07_09GRB.eps,width=0.45\linewidth,angle=0}
 \hspace{0.08\linewidth}
\epsfig{figure=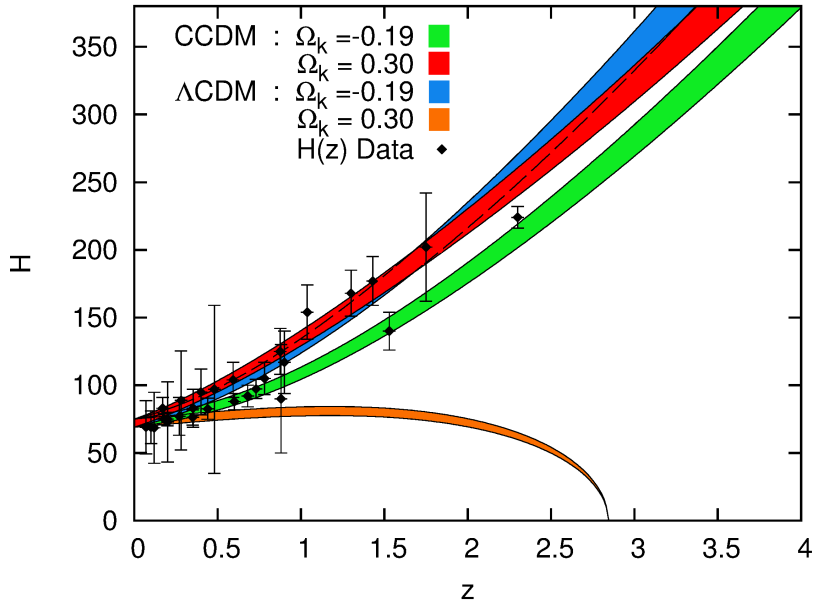,width=0.70\linewidth,angle=0}
}%-90}}height=3.8truein,
\caption{Comparison between $H(z)$ curves from CCDM and $\Lambda$CDM for some values of $\Omega_k$. The regions correspond to 68\% c.l. deviation over the $H_0$ value, as explained on the text. Also shown is the $H(z)$ data used for the statistical analysis.}
\label{hzcomparison}
\end{figure}

In fact, we have proposed a natural dependence of the creation rate over the expansion rate, and the direct inclusion of spatial curvature already breaks the ``dark degeneracy''. The question now is: can the Universe be nonflat enough in order to distinguish both models? Part of the answer is on Figure \ref{contours}. As we have seen, SNs+$H(z)$ alone, which are observational data quite independent from cosmological models, are not enough for constraining the spatial curvature. Namely, looking at its 95\% confidence limits over the curvature, we have $-0.19<\Omega_k<0.30$ so, the Universe can be flat, quite open ($\Omega_k\sim0.30$) or quite closed ($\Omega_k\sim-0.19$). Spatial curvature on the border of this limit is enough for distinguishing CCDM from $\Lambda$CDM, as can be seen on Figure \ref{hzcomparison}.

We have plotted, on Figure \ref{hzcomparison}, $H(z)$ curves for some values of $\Omega_k$, in the context of CCDM and $\Lambda$CDM. In all curves, we have fixed $\alpha=\Omega_\Lambda=0.792$, the CCDM best fit from SNs+$H(z)$, and we have used the $H_0$ value from \cite{Humphreys13}, $H_0=72.0\pm3.0$ kms$^{-1}$Mpc$^{-1}$, with the regions corresponding to 1-$\sigma$ variation over the best fit. As we can see, if the Universe is closed, CCDM and $\Lambda$CDM are harder to be distinguished, with a small distinction only at high redshifts. If, however, the Universe is open, the distinction can be made at intermediate redshifts, and, at high redshifts, the distinction is clear even with the current set of $H(z)$ data.

However, if we take into account the prior over $\Omega_k$ given by CMB, we can not distinguish them, as we have, in this case, an strict 95\% confidence limit over the curvature of $-0.104<\Omega_k<0.030$. This certainly is not enough to distinguish CCDM from $\Lambda$CDM. Here, one could think that CMB constraints on $\Lambda$CDM could not be used, directly, to constrain CCDM models, as the former model has not matter creation and the latter has. However, for the creation rate used here, $\Gamma\propto\frac{1}{H}$, creation is negligible on early Universe evolution, thus not changing the signatures imprinted on the last scattering surface. Thus, we conclude that the Universe is quite spatially flat, even in the context of this particular CCDM model, so we can not distinguish it from $\Lambda$CDM. The ``dark degeneracy'' is maintained from this analysis.

\section{\label{conclusion} Conclusion}

In this work we proposed a new cosmological model, based on the matter creation phenomena. We have shown that the proposed model is able to explain the background accelerating dynamics of the Universe, without a new dark fluid with negative pressure.

We have demonstrated that the present model is able to avoid the ``dark degeneracy'' simply through the presence of a baryonic content or the spatial curvature. For both cases, the model presents a  distinguishable Hubble expansion from the $\Lambda$CDM.

We have shown that this model can be distinguished from LJO with the inclusion of baryons and spatial curvature, including a relative difference of about 10$\%$ on $H(z)$, for $z\gtrsim 1$ and $\Omega_b\sim0.04$, as mentioned on Section V. 
Nevertheless, the current data from SNe Ia and $H(z)$ are still not able to observationally favor any of both models in the background level, as we have discussed.

Further investigations including perturbation analysis should be made, in order to realize if more significant   distinctions  between CCDM and $\Lambda$CDM models can be found at higher orders of density perturbations evolution.

\begin{acknowledgments}
The authors wish to thank J. A. S. Lima for very helpful discussions. J.F.J. is grateful to INCT-Astrof{\'i}sica and the Departamento de Astronomia (IAG-USP) for hospitality and facilities. FAO is supported by CNPq (Brazilian Research Agency).
\end{acknowledgments}

\end{document}